\pdfoutput=1
\documentclass[10pt,conference]{IEEEtran}
\IEEEoverridecommandlockouts

\usepackage{cite}
\usepackage{amsmath,amssymb,amsfonts}
\usepackage{algorithm}
\usepackage{algorithmic}

\usepackage{graphicx}
\usepackage{textcomp}
\usepackage{xcolor}
\usepackage{booktabs}
\usepackage{array}
\usepackage{listings}
\usepackage{microtype}
\usepackage{balance}
\usepackage{url}
\usepackage{tikz}
\usepackage{enumitem}

\usetikzlibrary{arrows.meta,positioning,shapes.geometric,fit,backgrounds,calc}
\usepackage{pgfplots}\pgfplotsset{compat=1.18}\usepgfplotslibrary{groupplots}
\usepackage[caption=false,font=footnotesize]{subfig}

\definecolor{mplblue}{RGB}{31,119,180}
\definecolor{mplorange}{RGB}{255,127,14}
\definecolor{mplgreen}{RGB}{44,160,44}
\pgfplotsset{rqax/.style={width=\linewidth, height=3.7cm,
  xlabel={decoding temp.}, xtick={0,0.3,0.5,0.7,0.9}, xmin=-0.05, xmax=0.95,
  tick label style={font=\tiny}, label style={font=\tiny},
  title style={font=\scriptsize}, legend cell align={left},
  legend style={font=\tiny, draw=none, fill=none, inner sep=1pt},
  every axis plot/.append style={thick, mark size=1.5pt}}}
\newcommand{\legbox}[1]{\tikz[baseline=-0.5ex]{\node[draw,rounded corners=1pt,minimum width=2.6mm,minimum height=2.6mm,inner sep=0pt,#1]{};}}

\newcommand{\secRef}[1]{\mbox{\S\,\ref{#1}}}

\lstdefinestyle{telemetry}{
  basicstyle=\ttfamily\scriptsize,
  breaklines=true,
  frame=single,
  framesep=4pt,
  xleftmargin=4pt,
  xrightmargin=4pt,
  columns=fullflexible,
  keepspaces=true,
}

\begin{document}

\title{ORCA: Observability-Grounded Program Repair for Microservice Incidents}

\author{%
\IEEEauthorblockN{%
Yuanchen Gao\textsuperscript{1,*},
Yifang Tian\textsuperscript{2},
Yiran Li\textsuperscript{2},
Charles Zhang\textsuperscript{1},
Hans-Arno Jacobsen\textsuperscript{2}%
\thanks{\textsuperscript{*}This work was conducted while Yuanchen Gao was an International Visiting Graduate Student (IVGS) at the University of Toronto.}}
\IEEEauthorblockA{\textsuperscript{1}Hong Kong University of Science and Technology\\
\{ygaocv@connect, charlesz@cse\}.ust.hk}
\IEEEauthorblockA{\textsuperscript{2}University of Toronto\\
yifang.tian@mail.utoronto.ca, one.li@utoronto.ca, jacobsen@eecg.toronto.edu}
}
\maketitle
\begin{abstract}
Microservice failures are often diagnosed from operational telemetry. However, automated program repair systems usually start from issue reports, localized code context, or failing tests. This mismatch leaves a gap between telemetry-based diagnosis and patch generation. We present \textsc{ORCA}, an observability-grounded APR pipeline for microservice incidents. \textsc{ORCA} first distills the differences in paired failure and reference telemetry into a fault signature, then uses the signature to identify candidate code and deployment-configuration locations. Repair graph agents and an Exploration agent generate unified-diff patch candidates from these locations.

\textsc{ORCA} evaluates generated patches with a Telemetry-Grounded Patch Verifier that separates patch validity, syntactic and semantic correctness, test-oracle integrity, and telemetry replay. On a 575-case benchmark, \textsc{ORCA} outperforms all evaluated baselines in terms of cost-effectiveness. Results show that operational telemetry can be transformed from diagnostic evidence into actionable repair context: paired telemetry supports repair-oriented localization, while repair graph
agents convert localized code and configuration evidence into constrained
patch-generation context for the LLM. Telemetry-grounded verification then
exposes repair outcomes that issue- or test-only evaluation would miss.

\end{abstract}


\section{Introduction}
\label{sec:intro}
Modern microservice applications are especially prone to failures that
arise from interactions among many independently deployed services,
configuration files, and runtime dependencies~\cite{zhou2021trainticket,
gan2019deathstarbench, opentelemetrydemo, googlecloudmicroservicesdemo}. In production, such failures are typically observed
through telemetry rather than through a failing test: logs expose exceptions
and abnormal messages, traces identify service operations on delayed or failed
request paths, and metrics capture shifts in latency, error rate, traffic
volume, and resource use~\cite{li2020microRCA,lee2023eadro,yu2023nezha,
pham2025rcaeval}. Although these signals support incident diagnosis, they do
not directly specify a repair. A telemetry anomaly may identify a degraded
service or request path~\cite{yu2021microrank,zhang2024tracecontrast}, while
the actual fix may require editing an application method, a deployment
manifest, an environment variable, or a configuration. The challenge is to map
heterogeneous, high-volume behavioral evidence to repository locations and to a
validation signal for a candidate patch.

Recent automated program repair (APR) systems are commonly developed and evaluated in issue-to-patch
settings.
In SWE-bench, for example, each task provides a
repository snapshot and a natural-language GitHub issue, and success is
measured by test outcomes~\cite{jimenez2024swebench}. Systems developed for
this setting therefore address issue-grounded repository editing, rather than repairs derived from paired telemetry~\cite{yang2024sweagent,
xia2024agentless,zhang2024autocoderover}.

Spectrum-based and learned fault localizers similarly rely on test
outcomes or explicit failure descriptions~\cite{abreu2007ochiai,
  kang2024autofl}, while classic APR validates candidate patches
against a test suite~\cite{legoues2012genprog,
  long2016prophet}. Microservice root-cause analysis (RCA) systems take the opposite
approach: they consume telemetry and rank suspicious services or call
paths~\cite{yu2021microrank, li2020microRCA}, but they stop at
diagnosis and do not produce repository patches. Configuration repair tools address deployment configuration
files~\cite{saavedra2025infrafix}, but usually analyze those files apart
from the log, trace, and metric evidence observed during the incident. As a
result, producing a repair patch from telemetry still requires
identifying candidate code or configuration locations and defining a
validation signal for the observed failure.

We propose \textsc{ORCA}, an observability-grounded automated program repair
pipeline for microservice incidents. \textsc{ORCA} takes paired failure and
reference telemetry bundles, the project workspace when available, and an
optional issue title and body. For real-incident cases, \textsc{ORCA}
sanitizes the issue title and body to reduce solution leakage by removing
explicit references to the upstream repair; we refer to the sanitized result as the sanitized issue context. \textsc{ORCA} emits a unified-diff patch candidate for
application code or deployment configuration.

\begin{figure*}[t]
\centering
\includegraphics[width=\textwidth]{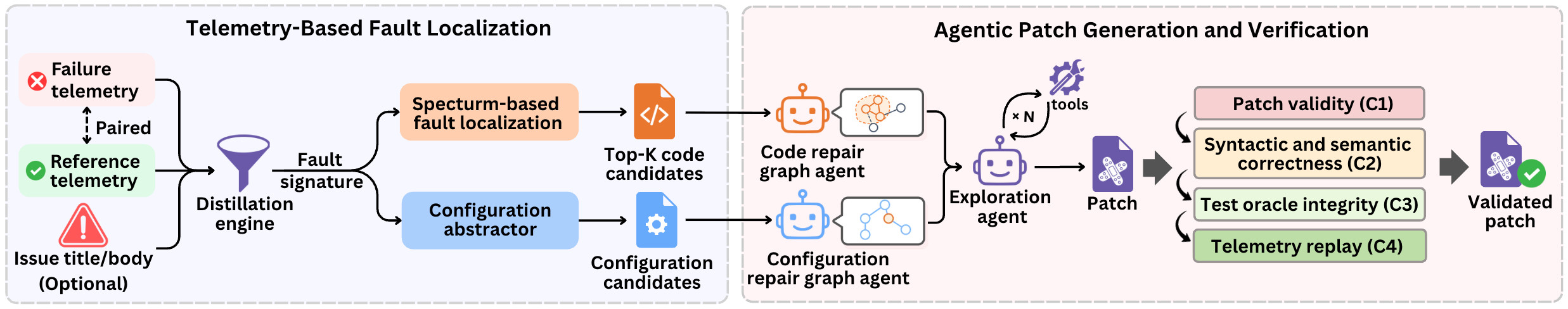}
\vspace{-15pt}
\caption{End-to-end \textsc{ORCA} pipeline}
\label{fig:pipeline}
\end{figure*}

\textsc{ORCA} has two stages, as shown in Figure~\ref{fig:pipeline}. The first stage, \textit{Telemetry-Based Fault
Localization}, compares failure and reference telemetry. The Distillation
engine summarizes telemetry differences as a compact \emph{fault signature}.
The Spectrum-based fault localization (SBFL) module uses the fault signature to
rank top-$K$ code candidates, while the Configuration abstractor selects
configuration candidates for deployment or telemetry faults. The second
stage, \textit{Agentic Patch Generation and Verification}, converts these
localized candidates into a patch candidate. It uses a code repair graph
agent for code candidates, a configuration repair graph agent for
configuration candidates, and the Exploration agent when repair graph
generation does not produce an accepted patch. The resulting patch candidate
is evaluated by the \textit{Telemetry-Grounded Patch Verifier} (TGPV), which
reports patch validity, syntactic and semantic correctness, test-oracle
integrity, and telemetry replay.

We evaluate \textsc{ORCA} on a 575-case benchmark spanning synthetic code,
synthetic configuration, and real microservice incidents, against six
representative baselines including issue-driven agentic APR systems. On
the synthetic code and configuration subsets, \textsc{ORCA} obtains the
highest counts for the reported verification checks among the evaluated systems. On the 150-case
real-incident subset, \textsc{ORCA} produces $113$ patches that pass patch
validity and $45$ telemetry-replay mitigations. The evaluated
baseline with the highest telemetry-replay-mitigation count, mini-SWE-agent~\cite{minisweagent2025}, obtains $74$ and
$23$, respectively, while using about $640$k mean tokens per case compared
with \textsc{ORCA}'s about $26$k.

\textsc{ORCA} makes three contributions:
\begin{enumerate}[leftmargin=*]
\item A telemetry-to-repair formulation for microservice incidents. The
formulation starts from paired failure and reference telemetry bundles,
distills their behavioral differences into a compact fault signature, and
uses the resulting telemetry evidence to derive candidate code and
configuration locations.

\item A two-stage repair pipeline for observability-grounded APR. The first
stage, Telemetry-Based Fault Localization, combines Spectrum-based fault
localization with a Configuration abstractor. The second stage, Agentic
Patch Generation and Verification, uses a code repair graph agent, a
configuration repair graph agent, and the fallback Exploration agent to produce a
unified-diff patch candidate under exact-match edit constraints and TGPV
verification.

\item A telemetry-grounded patch verification and evaluation procedure. ORCA
checks each generated patch along four axes: patch validity, syntactic and
semantic correctness, test-oracle integrity, and telemetry replay. The
evaluation uses these checks to compare \textsc{ORCA} and six baselines on a
575-case benchmark covering synthetic code faults, synthetic configuration
faults, and real microservice incidents.
\end{enumerate}

The remainder of this paper is organized as follows.
\secRef{sec:related} positions \textsc{ORCA} against existing
program-repair, RCA, and current agentic work.
\secRef{sec:approach} and \secRef{sec:hybrid} then present \textsc{ORCA}'s two-stage pipeline.
\secRef{sec:eval-setup} describes the datasets, baselines, metrics, and
experimental results. \secRef{sec:discussion} discusses
limitations, and \secRef{sec:concl} concludes.

\section{Related Work}
\label{sec:related}

\providecommand{\cmark}{\textcolor{green!50!black}{\checkmark}}
\providecommand{\xmark}{\textcolor{black!40}{\ensuremath{\times}}}

Table~\ref{tab:related-matrix} compares \textsc{ORCA} with representative
work from four areas: APR, fault localization,
microservice RCA, and configuration diagnosis or repair.
These areas address adjacent parts of observability-grounded repair, but
none covers the full telemetry-to-patch workflow setting. APR systems generate
patches, but typically assume a natural-language issue, a scoped code
context, or a failing test. Fault-localization techniques rank suspicious
program elements, but usually rely on test outcomes or explicit failure
descriptions. Microservice RCA methods consume logs, traces, and metrics,
but generally stop at service- or path-level diagnosis rather than
producing editable code or configuration changes. Configuration tools
address deployment configuration files, but usually analyze those files
apart from incident telemetry. \textsc{ORCA} targets the gap between
observability-based diagnosis and end-to-end repair for microservice
incidents.

\begin{table}[t]
  \caption{Capability comparison of \textsc{ORCA} with prior work.}
\label{tab:related-matrix}
\centering
\setlength{\tabcolsep}{3pt}\renewcommand{\arraystretch}{1.12}\scriptsize
\newcommand{\rh}[2]{\shortstack[c]{#1\\#2}}
\resizebox{\columnwidth}{!}{%
\begin{tabular}{@{}l cccccc@{}}
\toprule
\textbf{System / family} &
\rh{Telemetry-}{grounded} & \rh{Code}{FL} & \rh{Config}{FL} &
\rh{Code}{repair} & \rh{Config}{repair} & \rh{Multi-}{language} \\
\midrule
SWE-agent~\cite{yang2024sweagent,minisweagent2025} & \xmark & \cmark & \cmark$^{\ddagger}$ & \cmark & \cmark$^{\ddagger}$ & \cmark \\
ReAct / single-shot~\cite{yao2023react} & \xmark & \xmark & \xmark & \cmark & \cmark$^{\ddagger}$ & \cmark \\
Agentless~\cite{xia2024agentless} & \xmark & \cmark & \xmark & \cmark & \xmark & \xmark \\
AutoCodeRover~\cite{zhang2024autocoderover} & \xmark & \cmark & \xmark & \cmark & \xmark & \xmark \\
GenProg, Prophet~\cite{legoues2012genprog,long2016prophet} & \xmark & \cmark & \xmark & \cmark & \xmark & \xmark \\
ChatRepair~\cite{xia2024chatrepair} & \xmark & \xmark & \xmark & \cmark & \xmark & \cmark \\
SBFL~\cite{abreu2007ochiai,sohn2017fluccs} & \xmark & \cmark & \xmark & \xmark & \xmark & \cmark \\
Microservice RCA~\cite{yu2021microrank,yu2023nezha} & \cmark & \xmark & \xmark & \xmark & \xmark & \cmark \\
Config diagnosis~\cite{xu2013spex,zhang2013confdiagnoser,zhou2023mmd} & \xmark & \xmark & \cmark & \xmark & \xmark & \xmark \\
GLITCH~\cite{saavedra2022glitch} & \xmark & \xmark & \cmark & \xmark & \xmark & \xmark \\
InfraFix~\cite{saavedra2025infrafix} & \xmark & \xmark & \cmark & \xmark & \cmark & \xmark \\
\midrule
\textbf{\textsc{ORCA} (ours)} & \cmark & \cmark & \cmark & \cmark & \cmark & \cmark \\
\bottomrule
\end{tabular}}
\par\smallskip
{\scriptsize\raggedright
$\cmark^\ddagger$ indicates the work can inspect and
edit non-code files but can not perform dedicated configuration fault-localization or deployment-configuration repair
component.}
\end{table}

\noindent
\textbf{Automated program repair.}
Recent APR systems are often developed and evaluated in issue-to-patch
settings. SWE-bench provides a repository snapshot and a natural-language
GitHub issue, and measures success with test
outcomes~\cite{jimenez2024swebench}; SWE-agent implements an interactive
repository-editing agent for this setting~\cite{yang2024sweagent,
minisweagent2025}. Agentless performs staged localization from files to
elements and lines before repair and uses a SEARCH/REPLACE patch
format~\cite{xia2024agentless}, while
AutoCodeRover~\cite{zhang2024autocoderover} combines code search with
localization to guide repository editing. Conversational repair with
compiler and test feedback~\cite{xia2024chatrepair} and earlier APR
methods~\cite{xia2023llmapr,legoues2012genprog,long2016prophet} further
establish issue- and test-based APR. These systems are designed around
natural-language issue reports, code context, or executable tests, rather than paired
failure and reference telemetry. \textsc{ORCA} addresses a different input
setting: it starts from paired telemetry bundles and uses telemetry-derived localization to guide and constrain patch generation. It also reports separate patch validity,
syntactic and semantic correctness, test-oracle integrity, and telemetry
replay outcomes rather than a single test-suite result.

\noindent
\textbf{Fault localization.}
SBFL ranks program elements by
comparing their coverage in failing and passing
tests~\cite{jones2005tarantula,abreu2007ochiai}. Subsequent work augments
SBFL with code and change metrics~\cite{sohn2017fluccs}, learns feature
combinations with deep models~\cite{li2019deepfl}, or uses LLMs to
generate fault locations and explanations~\cite{kang2024autofl}. These
techniques localize faults within code, but their evidence is usually a
test outcome or an explicit failure description. \textsc{ORCA} uses the
same suspiciousness principle with paired failure and reference telemetry: log
templates, trace operations, and metric series provide the observations
used to score code and configuration candidates.

\noindent
\textbf{Microservice RCA.}
Microservice RCA methods use telemetry to identify the service or request
path most likely to explain an incident~\cite{zhang2024microservicesurvey,zhanglongAIOPS2025}. Metric-based methods model
service dependencies and propagate anomaly scores over those
dependencies~\cite{li2020microRCA,wang2018cloudranger,pham2025rcaeval},
while trace-based methods analyze distributed traces to identify anomalous
operations or causal paths~\cite{liu2020traceanomaly,yu2023nezha,
lee2023eadro}. These methods are telemetry-grounded, but their output is
diagnostic: a suspect service, operation, or request path. \textsc{ORCA} uses related telemetry signals
for repair-oriented localization. Its scoring incorporates
frequency-aware trace evidence, critical-path latency evidence, and metric
anomaly evidence inspired by MicroRank~\cite{yu2021microrank}, TraceContrast~\cite{zhang2024tracecontrast}, and
Baro~\cite{pham2024baro}, but maps
the resulting evidence to code and configuration locations and then
generates a patch.

\noindent
\textbf{Configuration diagnosis and repair.}
Configuration diagnosis includes analyses that detect misconfigurations
from inferred constraints or from differences between failing and passing
profiles~\cite{xu2013spex,xu2016pcheck,zhang2013confdiagnoser,
attariyan2010confaid}, as well as cross-file analyses that discover
correlated configuration dependencies~\cite{zhou2023mmd,chen2020cdep,
mahmud2023confd}. Repair-oriented work detects or repairs defects in
infrastructure-as-code and deployment configuration
files~\cite{saavedra2025infrafix,saavedra2022glitch}, validates
configuration files with LLMs~\cite{lian2025ciri,wen2026slsdetector}, or
checks configuration changes for regressions~\cite{chen2025stratus}.
These systems usually inspect configuration files directly, rather than
connecting them to the log, trace, and metric evidence observed during an
incident.
\textsc{ORCA} differs by selecting configuration candidates from incident
telemetry and repairing related deployment files within the same
patch-generation pipeline.

\section{Telemetry-Based Fault Localization}
\label{sec:approach}
\label{sec:distill}
\label{sec:fl-stage}
Microservice telemetry is valuable for diagnosing production failures,
but it does not identify repair locations directly. Logs, traces, and
metrics expose abnormal behavior at the level of services, operations,
and runtime measurements; a patch, however, must modify a concrete
method or some configuration file. \textsc{ORCA}'s telemetry-based
\textit{fault localization} addresses this challenge without invoking a
language model. It first compares failure and reference telemetry and
summarizes their differences as a compact \emph{fault signature}. It
then scores candidate code and configuration locations from that
signature: methods for source-code faults and configuration files for
deployment or telemetry faults. The resulting candidates provide
the patch-generation stage with scoped, telemetry-derived edit locations.

\noindent \textbf{Distillation engine.}
The fault signature is a structured textual summary of the differences
between the failure telemetry and the reference
telemetry. \textsc{ORCA} groups each telemetry source and orders the
groups by deterministic failure-reference comparison rules. For
logs, \textsc{ORCA} replaces volatile fields such as timestamps,
request identifiers, memory addresses, and hash values with
placeholders, groups records by severity and message template, and
ranks templates by severity and by the difference between their
failure-side and reference-side counts. For traces, \textsc{ORCA}
groups spans by service and operation, and ranks each group by three
quantities: the failure-reference call-count difference, the latency
increase relative to the reference side, and the number of
error-status spans. For metrics, \textsc{ORCA} ranks service-level
series by their deviation from the reference window, including CPU
usage, memory usage, request latency, and throughput. \textsc{ORCA}
applies fixed caps per telemetry source across deployment substrates.

The localization step adapts spectrum-based fault
localization~\cite{abreu2007ochiai,sohn2017fluccs} from test coverage
to failure-reference telemetry. It defines two candidate types. A
\emph{code candidate} is a method in a source file. A
\emph{configuration candidate} is a configuration file involved in deployment,
runtime, or build behavior, such as Dockerfiles, Docker Compose files,
Kubernetes YAML, \texttt{.env} files, collector configurations, shell scripts,
or Makefiles. \textsc{ORCA} scores code candidates and selects configuration
candidates separately, as described below.

\noindent \textbf{Code-candidate scoring.}
\textsc{ORCA} scores each code candidate from three telemetry signals: logs, traces, and metrics. For each signal $X\in\{L,T,M\}$, it computes an Ochiai-based suspiciousness
score $s_X(c)$ for candidate $c$~\cite{abreu2007ochiai}:
\[
s_X(c)=
\frac{\epsilon_{ef}^{X}(c)}
{\sqrt{(\epsilon_{ef}^{X}(c)+\epsilon_{nf}^{X}(c))(\epsilon_{ef}^{X}(c)+\epsilon_{ep}^{X}(c))}}.
\]
Here, $\epsilon_{ef}^{X}(c)$ denotes failure-side telemetry evidence
from signal $X$ attributable to $c$, $\epsilon_{nf}^{X}(c)$ denotes
failure-side telemetry evidence from signal $X$ not attributable to
$c$, and $\epsilon_{ep}^{X}(c)$ denotes reference-side telemetry
evidence from signal $X$ attributable to $c$. The score increases when
telemetry evidence for $c$ is concentrated on the failure side and is
rare or absent on the reference side. Unlike traditional SBFL, which
applies Ochiai to test
coverage~\cite{jones2005tarantula}, \textsc{ORCA} applies it to
telemetry observations, following the code-metric extension of
FLUCCS~\cite{sohn2017fluccs}.

\emph{Log signal} treats each severity-ranked message template as an observation. A
template associated with candidate $c$ contributes to
$\epsilon_{ef}^{L}(c)$ when it appears on the failure side and to
$\epsilon_{ep}^{L}(c)$ when it appears on the reference
side. Templates with higher severity and larger failure-reference
count differences therefore contribute more fault evidence.

\emph{Trace signal}
scores service-operation groups and adds two refinements to membership
counts. First, it weights each trace operation $o$ by the count
difference $\Delta(o)=\max(0,n_f(o)-n_r(o))$, where $n_f(o)$ and
$n_r(o)$ are the failure-side and reference-side counts of
$o$~\cite{yu2021microrank}. This assigns more evidence to operations
whose frequency increases on the failure side than to operations that
are common in both runs. Second, for each trace, \textsc{ORCA}
computes the span chain with the largest cumulative duration and adds
failure-side evidence for candidates on that chain. This captures
latency evidence even when an operation appears on both sides of the
paired telemetry.

\emph{Metric signal}
first ranks CPU, memory, latency, and throughput series by their
standardized deviation from the reference window and tests each series
for a mean shift beyond the change-point threshold. Operation-tagged
series, such as the latency or throughput of a specific endpoint, are
attributed to the method implementing that endpoint; series without
operation tags are attributed to candidate methods in the affected
service. For each attributed series, the contribution to
$\epsilon_{ef}^{M}(c)$ is the standardized deviation from the
reference distribution,
$z(v)=|v-\mu_{\text{ref}}|/\sigma_{\text{ref}}$, capped at $5$ in the
implementation to limit the influence of extreme outliers. If the same
series also exhibits a mean shift, \textsc{ORCA} adds one additional
failure-side observation to the attributed candidate.

The final code-candidate score is the unweighted mean
$\mathit{final}(c)=(s_L(c)+s_T(c)+s_M(c))/3$. This aggregation keeps
the three telemetry sources on the same Ochiai-normalized scale: logs
provide message-level failure evidence, traces provide
operation-frequency and latency evidence along request paths, and
metrics provide service-level resource and request-behavior
evidence. \textsc{ORCA} returns the top-$K$ code candidates, with $K=10$ in our implementation.

\noindent \textbf{Configuration abstractor.}
\label{sec:fl-config}
Code candidates do not cover faults whose required edit is a
Dockerfile, a Kubernetes manifest, a build script, or a collector
configuration entry. The Configuration abstractor handles these
cases with an independent pass over configuration files. \textsc{ORCA} considers configuration candidates only when the fault
signature suggests a deployment or instrumentation fault, optionally
supported by the sanitized issue context, rather than a source-code fault. We use two conditions for this decision: \textit{deployment
instability}, where workloads are not ready and telemetry streams are
missing or incomplete, and \textit{series-set expansion}, where the
failure run produces many more distinct metric labels or trace
operation names than the reference run. If neither condition
holds, \textsc{ORCA} returns no configuration candidates. Otherwise,
it selects all configuration files whose owning service or file type
matches the fault signature or, when available, the sanitized issue context. The ranked code candidates and the selected configuration
candidates are passed to the \textit{Agentic Patch Generation and
Verification} stage.

\section{Agentic Patch Generation and Verification}

\label{sec:hybrid}
The second stage of \textsc{ORCA} turns telemetry-localized evidence into a patch candidate and then performs telemetry-grounded patch verification (TGPV). Patch generation is agentic because a valid edit may require repository exploration, multi-file reasoning, and language-model generation. However, unlike issue-driven agents such as SWE-agent~\cite{yang2024sweagent} and ReAct~\cite{yao2023react}, \textsc{ORCA} does not start from an unconstrained natural-language prompt. Each model invocation is grounded in the fault signature, the localized code or configuration candidates, and a one-hop repair graph over candidate files and methods.

A deterministic dispatcher maps the localization outputs to the patch-generation
agents in Algorithm~\ref{alg:dispatch}. \textsc{ORCA} has two repair graph
agents, one for code candidates and one for configuration candidates. These agents are not mutually exclusive: an incident can require both application-code and deployment-configuration edits, in which case \textsc{ORCA} invokes both repair graph agents and merges their patch candidates into a single patch.

\begin{algorithm}[t]
\caption{\textsc{ORCA} patch-generation dispatch and verification. Code and configuration repair graph agents use specific prompts, and both may be selected before TGPV verification.}
\label{alg:dispatch}
\small
\begin{algorithmic}[1]
\REQUIRE $\sigma$ fault signature; $\mathcal{B}$ ranked code-candidate list with top suspiciousness score $s_{\max}$; $C$ configuration candidate set; code repair graph agent threshold $\tau$
\ENSURE patch candidate $p$ and verifier outcomes $v$
\STATE $\mathcal{P} \leftarrow \emptyset$
\STATE \textbf{if} $C \neq \emptyset$ \textbf{then} $\mathcal{P} \leftarrow \mathcal{P} \cup \textsc{ConfigRepairGraphAgent}(C,\sigma)$
\STATE \textbf{if} $s_{\max} \geq \tau$ \textbf{then} $\mathcal{P} \leftarrow \mathcal{P} \cup \textsc{CodeRepairGraphAgent}(\mathcal{B},\sigma)$
\STATE \textbf{if} $\mathcal{P} = \emptyset$ \textbf{then} $\mathcal{P} \leftarrow \textsc{ExplorationAgent}(\sigma,\mathcal{B},C)$
\STATE $p \leftarrow \textsc{Merge}(\mathcal{P})$
\STATE \textbf{if} $p \neq \emptyset$ \textbf{then} $v \leftarrow \textsc{TGPV}(p,\sigma)$ \textbf{else} $v \leftarrow \emptyset$
\STATE \textbf{return} $(p,v)$
\end{algorithmic}
\end{algorithm}

The dispatcher routes cases using the localization outputs. Line~2 invokes the
configuration repair graph agent when the configuration candidate set $C$ is
nonempty. Line~3 invokes the code repair graph agent when the highest
code-candidate suspiciousness score $s_{\max}$ satisfies
$s_{\max}\geq\tau$; we set $\tau=0.60$ and keep it fixed across deployment
environments. The two repair graph agents are constrained LLM calls, described
below, and may both contribute patch candidates for the same incident. If no
repair graph agent produces a candidate, Line~4 invokes the Exploration agent
with the fault signature, ranked code candidates, and configuration
candidates. \textsc{ORCA} launches $N$ independent Exploration-agent runs and combines their outputs with file-level majority voting. Line~5
merges the selected candidates into a unified-diff patch candidate; if no agent
produces a patch, \textsc{Merge} returns the empty patch. Line~6 applies TGPV
only when the merged patch is nonempty and otherwise returns empty verifier
outcomes. Line~7 returns the patch and verifier outcomes.

\noindent \textbf{Repair graph agents.} Each repair graph agent invoked
by the dispatcher issues one LLM call using a specific prompt. The code
and configuration agents share the same high-level prompt structure, but differ
in their edit context, repair graph summary, and repair instructions. Each
prompt contains five ordered parts: failure-side evidence from the fault
signature, a structured edit context derived from localization, target-file
context, a textual repair graph summary, and an exact-match edit instruction.
The structured edit context is a JSON-formatted object that packages the
localization output for the corresponding agent. For code repair, it contains
the fault signature $\sigma$, the ranked method candidates $\mathcal{B}$ with
their suspiciousness scores, the affected service when it can be inferred from
telemetry, and a repository reference. When a code workspace is available, the
tool resolves each candidate to a concrete \textit{file:line} location;
synthetic bundles include this mapping from the injector. For configuration
repair, the edit context contains the configuration candidate files in $C$
rather than method candidates. In both repair graph agents, the context defines
the target locations the model may edit and the telemetry evidence associated
with those locations.

The model must return an exact-match search-and-replace edit: the search text must be copied verbatim from the target file, and the replacement text specifies the new content for that region. The agent accepts the output only when the search text matches the target file exactly; otherwise it rejects the edit before verifier execution. Accepted edits are converted into the unified-diff patch candidate returned by the repair agent.

\begin{figure*}[t]
\centering
\subfloat[Code repair graph]{%
\includegraphics[width=0.49\textwidth]{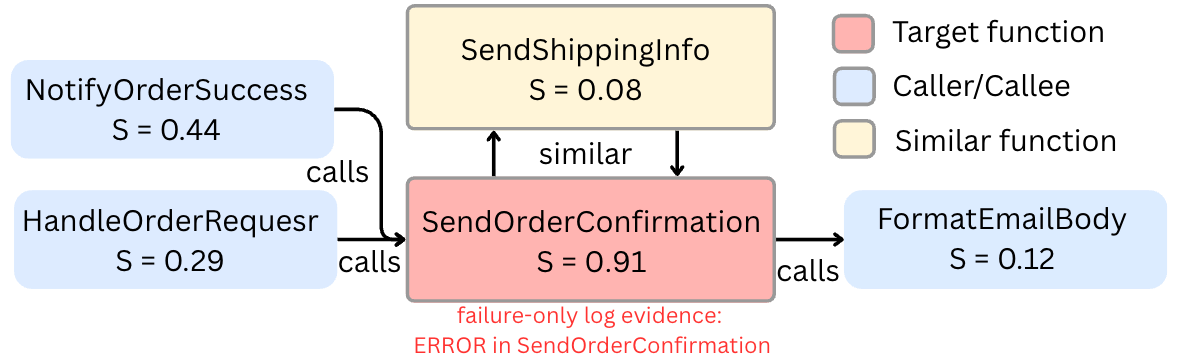}%
\label{fig:repair-graph}}
\hfil
\subfloat[Configuration repair graph]{%
\includegraphics[width=0.49\textwidth]{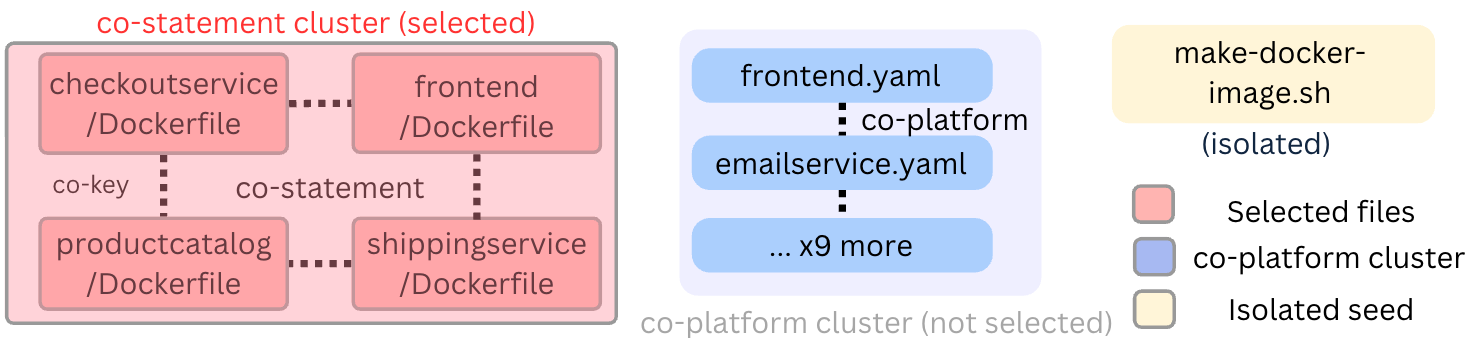}%
\label{fig:config-repair-graph}}
\vspace{-5pt}
\caption{Illustrations of code (a synthetic Python case) and configuration (an example from \textit{microservices-demo}~\cite{googlecloudmicroservicesdemo}) repair graphs.}
\label{fig:graphs}
\end{figure*}

\noindent{\textbf{Code repair graph.}} Localization gives the repair agent a ranked list of suspicious methods, but the correct edit may depend on the surrounding call context. The code repair graph therefore augments the top-ranked candidate with one-hop structural context: callers, callees, and textually similar methods, each annotated with its telemetry-derived suspiciousness score. This context helps the model distinguish whether the fault is in the localized method itself, in a caller that passes incorrect inputs, or in a callee whose behavior must be adjusted. The prompt serializes this graph as a short textual summary listing each neighboring method, its score, and its relation to the target. Figure~\ref{fig:graphs}(a) illustrates this context for an email-service fault, where the failure evidence is concentrated on \textit{SendOrderConfirmation} while related methods provide the edit boundary.

\noindent{\textbf{Configuration repair graph.}} Configuration localization returns a set of candidate files rather than a single ranked method, and a correct configuration fix may require coordinated edits across related files. The configuration repair graph groups candidate files that share structural or semantic configuration context, such as the same service, file type, deployment directory, or configuration key. \textsc{ORCA} then selects the group whose file type and service context are consistent with the fault signature and sends all files in that group to one model call. This graph context is needed because configuration faults often span files that do not individually expose runtime divergence but must be edited together. Figure~\ref{fig:graphs}(b) shows such a case: the selected co-statement cluster contains multiple service configuration files, while an isolated build script and a co-platform cluster are not selected.

\noindent \textbf{Exploration agent.}
\label{sec:hybrid-exploration}
The Exploration agent is the fallback patch-generation path. It is selected
when neither repair graph agent produces an accepted exact-match patch. This
fallback covers faults whose repair location is not recovered by the ranked
code or configuration candidates, including semantic faults, naming or
convention faults, and build faults outside those candidate sets.

The Exploration agent is an interactive repository-level repair agent. Each
agent instance operates in a sandboxed repository workspace, with an action
space that includes file inspection, repository search, file editing, and
validation commands. An instance either produces repository edits that are
converted into a unified-diff patch candidate or abstains when it cannot
identify a repair location. An empty-edit check rejects outputs with no
repository change and treats step-budget exhaustion without edits as an
abstention.

\textsc{ORCA} can run $N$ independent Exploration-agent instances for the same
case. When $N>1$, it combines their outputs with file-level majority voting:
the selected patch is taken from the instance that edited the file selected by
the largest number of instances. Unless otherwise stated, the main experiments
use $N{=}1$ and an interaction-step budget of $B{=}15$ per instance.

For configuration-oriented repairs, the Exploration agent can request a bounded
set of related configuration files from the configuration repair graph. This
allows a local configuration edit to be considered together with files that may
require the same change.

\noindent \textbf{Telemetry-Grounded Patch Verifier.} Given a patch candidate produced by the selected repair agents or the exploration fallback, TGPV verifies whether the patch is usable and whether it mitigates the telemetry fault. A single pass/fail signal is insufficient for microservice repair: a patch may apply but fail to deploy, pass tests only because it weakens the test oracle, or change code unrelated to the request path that triggered the incident. TGPV therefore separates verification into four checks.

\emph{Patch validity (\textbf{C1}).} This check determines whether the generated patch applies to the workspace. TGPV first tries the patch as generated; if it fails, TGPV normalizes common LLM artifacts such as Markdown code-block delimiters and workspace-relative file paths, and then retries with limited fuzz tolerance.

\emph{Syntactic and semantic correctness (\textbf{C2}).} This check is applied after patch validity and evaluates the patch against the
correctness criteria provided by the benchmark.
In the synthetic code subset, Java instances are checked with their paired
fault-revealing unit tests. Python instances are checked against the seeded
mutation and the reference pre-injection version; a patch passes C2 when it
restores the mutated method with respect to that reference. Accordingly, C2 is
reported only for the synthetic code subset, where the benchmark provides these
case-specific checks.

\emph{Test-oracle integrity (\textbf{C3}).} This check rejects patches that obtain an apparent test pass by corrupting the test oracle. It flags patches that edit test files, delete \textit{@Test} annotations, or weaken assertions.

\emph{Telemetry replay (\textbf{C4}).} This check redeploys the patched stack, replays the same traffic profile, and recomputes the telemetry signal associated with the failure-side fault signature. Telemetry replay is used as a diagnostic signal rather than a complete correctness oracle, so TGPV reports it at two thresholds. \emph{C4-strict} requires the failure-side signature to drop below $10\%$ of its failure-run baseline while expected-behavior indicators remain above their thresholds. \emph{C4-mitigated} requires the patch to apply, the workload to deploy, and the failure-side signature to decrease relative to the failure execution.

The four checks are reported separately because they capture different failure
modes: a patch may apply but fail the case-specific correctness check, preserve
the test oracle but leave replayed telemetry unchanged, or reduce the replayed
telemetry signal without editing an upstream repair file.

\section{Evaluation}
\label{sec:eval-setup}

We evaluate \textsc{ORCA} against the baselines and examine how model and
parameter choices affect the observed outcomes through the following research
questions:
\begin{description}\setlength{\itemsep}{0pt}
\item[\textbf{RQ1.}] How does \textsc{ORCA} compare against
six baselines on file-level localization and patch-verification outcomes
across code, configuration, and real-incident subsets?

\item[\textbf{RQ2.}] How sensitive are the results to the LLM backbone?
\item[\textbf{RQ3.}] How does the number of Exploration-agent runs
$N$ affect reported repair outcomes and token cost?
\item[\textbf{RQ4.}] How sensitive are the results to LLM temperature, i.e.,
the decoding temperature used during generation?
\end{description}

We also report a controlled ablation (Section~\ref{sec:ablation-telemetry})
that removes both the fault signature and the sanitized issue context derived
from the issue title and body, isolating their joint contribution.

\subsection{Datasets}
\label{sec:datasets}

Our benchmark contains 575 microservice failure cases from three sources and will be made publicly available. Each case is the unit of
evaluation, including a paired telemetry bundle collected from a failure
execution and a reference execution under the same workload and five-minute time window.

\emph{Synthetic code subset (200 cases).} This subset isolates the code-repair path. Following the controlled fault-injection model in the RCAEval benchmark~\cite{pham2025rcaeval}, the injector instantiates faults from twelve families across two widely used microservice systems, yielding 200 cases: six families in the Python code of Online Boutique (OB)~\cite{googlecloudmicroservicesdemo} and six in the Java code of Train-Ticket (TT)~\cite{zhou2021trainticket}. Because each fault is injected deterministically, the inverse of the injected edit provides the ground-truth repair for each case.

\emph{Synthetic configuration subset (225 cases).} This subset specifically addresses configuration repair. We construct $225$ controlled cases, each obtained by changing a single field in a single deployment resource; the reference repair is the inverse of the injected field change. The injected faults follow the categories of AIOpsLab~\cite{aiopslab2025}, spanning scheduling, resource limits, images, commands, probes, and routing. We apply these faults to four microservice testbeds: two applications in DeathStarBench~\cite{gan2019deathstarbench}, OpenTelemetry Demo~\cite{opentelemetrydemo}, and a TiDB operator deployment~\cite{pingcap_tidb_operator}.

\emph{Real-incident subset (150 cases).} This subset combines $56$ reproduced real incidents with $94$ controlled mutation variants derived from seed incidents. The $56$ reproduced incidents are drawn from fixed issues in four open-source repositories: \textit{microservices-demo}~\cite{googlecloudmicroservicesdemo}, \textit{opentelemetry-demo}~\cite{opentelemetrydemo}, \textit{train-ticket}~\cite{zhou2021trainticket}, and \textit{DeathStarBench}~\cite{gan2019deathstarbench}. For each issue, we identify the versions immediately before and after the repair and deploy both in containerized environments. We include a reproduced incident only when it satisfies four criteria: telemetry is available from both the faulty and fixed executions; the faulty revision has a materialized workspace; the ground-truth fix location contains between $1$ and $12$ substantive files; and the localization pool contains at least $100$ candidate snippets. The included incidents span eight source languages and fourteen ground-truth file types, including source files, Dockerfiles, shell scripts, Makefiles, docker-compose files, Kubernetes YAML, and \texttt{.env} files. The mutation portion adds $94$ controlled variants derived from nine real-incident seed cases by preserving the original fault pattern while changing the concrete program or configuration element that instantiates the fault, such as string constants, Boolean settings, numeric parameters, ports, timeouts, identifiers, or configuration keys. 

Before running any system, we sanitize each benchmark bundle to remove direct
repair hints. Removed fields include issue or pull-request identifiers,
upstream repair titles, ground-truth file paths, injected fault labels, and
per-snippet ground-truth labels.

\subsection{Baselines}
\label{sec:baselines}

We compare \textsc{ORCA} against six baselines, all run on the same
model at temperature $0$ and low reasoning effort. \emph{Direct} is a single LLM call that receives
the fault signature but no localized candidates or interaction loop.
\emph{One-Shot} is a single LLM call augmented with the top-five localized
candidates selected by
\textsc{ORCA}'s localization stage. \emph{ReAct}~\cite{yao2023react}
is a bounded tool-use loop with \textit{read\_file}, \textit{search},
\textit{list\_dir}, and \textit{submit\_patch} actions, capped at eight
steps. \emph{Agentless}~\cite{xia2024agentless} is the upstream
hierarchical localizer and repair pipeline, which refines the search from
files to program elements and then to lines before applying its
SEARCH/REPLACE post-processor. \emph{AutoCodeRover}~\cite{zhang2024autocoderover}
is a structure-aware repair agent that uses abstract syntax tree (AST)
code-search APIs with iterative context retrieval. Both released
implementations use Python-specific localization components; in the
synthetic code subset, we therefore run them only on the Python cases and, in the real-incident subset, treat them as Python-only baselines.
\emph{mini-SWE-agent}~\cite{minisweagent2025} is a shell-loop repair agent;
it edits files directly and emits \textit{git diff} on completion, using
the SWE-bench-style task prompt for which it was designed.
\emph{\textsc{ORCA}} is the full pipeline of Section~\ref{sec:approach}.
All baselines and \textsc{ORCA} receive the same fault signature and,
when the issue title and body are available, the same sanitized issue context.

\subsection{Metrics}
\label{sec:metrics}
We report effectiveness and efficiency metrics. Following
\secRef{sec:hybrid}, the tables and figures use C1--C4 to denote the four
TGPV checks. Effectiveness is reported primarily with patch validity (C1),
syntactic and semantic correctness (C2), and telemetry replay (C4). We
report test-oracle integrity (C3) separately to identify patches that edit
tests, remove test annotations, or weaken assertions instead of repairing
the fault. Efficiency is reported as average token cost and wall-clock time
per case. For real incidents, where exact patch equivalence is not generally
well-defined, we also report \textit{File hit}, a file-level localization
measure indicating whether the patch touches a file modified by the upstream
repair. All applicable baselines use the same metric definitions.

\noindent
\textbf{Primary real-incident metrics.}
For the synthetic subsets, injected faults provide known ground-truth files
or reference edits. For real incidents, exact textual patch matching is not
generally well-defined because a generated patch may address the same
incident through a different but semantically equivalent edit. We therefore
report File hit as only a file-level localization measure, use patch
validity to record whether the generated diff applies to the checked-out
workspace, and use telemetry replay as a separate runtime diagnostic.

Telemetry replay uses the C4-strict and C4-mitigated thresholds defined in
\secRef{sec:hybrid}. Cases without separable failure/fixed telemetry are
marked inconclusive and omitted from C4 totals. For cross-baseline telemetry
comparison, we use C4-mitigated and report C4-strict as the stricter
diagnostic. A patch may satisfy telemetry replay without a File hit when it changes the same fault-propagation path through a different file or location. Such
File-miss telemetry-replay outcomes require separate reporting and case-level
interpretation; Section~\ref{sec:res-c4} gives an example.

\subsection{RQ1: Cross-baseline comparison}
RQ1 evaluates \textsc{ORCA} and six baselines on the full 575-case
benchmark using DeepSeek-V4-pro~\cite{deepseek_v4_pro} with temperature $0$.

\subsubsection{Synthetic code subset (200 cases)}
\label{sec:res-syn}

\begin{table*}[t]
\renewcommand{\arraystretch}{1.2}
\caption{Results on synthetic subsets. Left: $200$ synthetic code faults. Right: $225$ synthetic configuration faults.}
\label{tab:syn200}
\centering\scriptsize\setlength{\tabcolsep}{3.2pt}
\vspace{-5pt}
\begin{tabular}{@{}lrr rr rrrr rrr @{\hskip 6pt} rrrrrr@{}}
\toprule
 & \multicolumn{11}{c}{\textbf{Synthetic code subset} ($200$)} & \multicolumn{6}{c}{\textbf{Synthetic config subset} ($225$)} \\
\cmidrule(lr){2-12}\cmidrule(l){13-18}
 & & & & & \multicolumn{4}{c}{OB-Python (100)} & \multicolumn{3}{c}{TT-Java (100)} & & & & & & \\
\cmidrule(lr){6-9}\cmidrule(lr){10-12}
\textbf{Method} & \textbf{Emitted patches} & \textbf{File hit} & \textbf{Tokens} & \textbf{Time (s)} & \textbf{C1} & \textbf{C2} & \textbf{C3} & \textbf{C4} & \textbf{C1} & \textbf{C2} & \textbf{C3} & \textbf{File hit} & \textbf{C1} & \textbf{C4s} & \textbf{C4m} & \textbf{Tokens} & \textbf{Time (s)} \\
\midrule
mini-SWE-agent & $63/200$ & $63/200$ & $23.1$k & $373$ & $34$ & $24$ & $\mathbf{100}$ & $29$ & $25$ & $17$ & $99$ & $177$ & $177$ & $110$ & $174$ & $113.4$k & $58.3$ \\
ReAct & $184/200$ & $177/200$ & $18.0$k & $149$ & $1$ & $1$ & $\mathbf{100}$ & $1$ & $4$ & $2$ & $99$ & $207$ & $30$ & $38$ & $57$ & $42.0$k & $77.7$ \\
Direct & $162/200$ & $160/200$ & $3.5$k & $65$ & $1$ & $1$ & $\mathbf{100}$ & $1$ & $0$ & $0$ & {$\mathbf{100}$} & $\mathbf{218}$ & $41$ & $24$ & $37$ & $6.2$k & $63.5$ \\
One-Shot & $168/200$ & $149/200$ & $3.7$k & $66$ & $5$ & $1$ & $\mathbf{100}$ & $2$ & $0$ & $0$ & $99$ & $215$ & $40$ & $21$ & $34$ & $5.0$k & $47.5$ \\
AutoCodeRover & $96/100$ & $96/100$ & $37.0$k & $221$ & $96$ & $56$ & $\mathbf{100}$ & $83$ & n/a & n/a & n/a & n/a & n/a & n/a & n/a & n/a & n/a \\
Agentless & $73/100$ & $73/100$ & $9.1$k & $155$ & $0$ & $0$ & $\mathbf{100}$ & $0$ & n/a & n/a & n/a & n/a & n/a & n/a & n/a & n/a & n/a \\
\midrule
\textbf{\textsc{ORCA}} & $\mathbf{197/200}$ & $\mathbf{193/200}$ & $34.0$k & $167$ & {$\mathbf{96}$} & {$\mathbf{58}$} & {$\mathbf{100}$} & {$\mathbf{95}$} & {$\mathbf{71}$} & {$\mathbf{30}$}& {$\mathbf{100}$} & $207$ & {$\mathbf{213}$} & {$\mathbf{131}$} & {$\mathbf{210}$} & $35.5$k & $93.5$ \\
\bottomrule
\end{tabular}
\par\smallskip
\makebox[\textwidth][c]{%
\scriptsize\emph{Notes:} C4s/C4m denote strict/mitigated telemetry replay.
Token consumption and time are per-case averages.}
\end{table*}

The left half of Table~\ref{tab:syn200} summarizes the synthetic code subset.
\textsc{ORCA} emits patches for $197$ of $200$ cases and obtains File hits in $193$ cases. On OB-Python, \textsc{ORCA} matches AutoCodeRover on
patch validity and obtains higher syntactic and semantic correctness and
telemetry replay counts. On TT-Java, where Agentless and
AutoCodeRover are not applicable, \textsc{ORCA} obtains the highest patch
validity and syntactic and semantic correctness counts among the
evaluated systems.

The comparison shows that a File hit does not necessarily correspond to a
verified repair. Direct, One-Shot, and ReAct often produce File hits, but few
of their emitted diffs pass the patch-validity check, which limits their
syntactic and semantic correctness and telemetry replay outcomes.
Agentless emits edits against paths that do not match the
injection workspace in these runs, yielding no diffs that pass the
patch-validity check. mini-SWE-agent is language-agnostic, but obtains lower
verification counts than \textsc{ORCA} on both OB-Python and TT-Java.
AutoCodeRover is competitive on OB-Python, but its released localization
components are Python-specific and it is not evaluated on TT-Java. Overall,
the synthetic code results indicate that \textsc{ORCA}'s advantage comes less
from file-level localization alone than from converting localized telemetry
evidence into patches that pass patch validity and satisfy the available
repair checks.

\noindent
\textbf{Dispatch and cost.}
The dispatcher sent $77$ of the $200$ cases to the repair graph agents,
where each case uses one bounded LLM call (about $10$k median tokens). The
remaining $123$ cases entered the Exploration agent: $103$ completed through
the default $N{=}1$ exploration run with the secondary tool-use loop, while
$20$ terminated after abstaining or producing an empty patch. This routing explains the two-level token-cost distribution: repair graph cases require one bounded
model call, whereas exploration-agent cases incur additional workspace-search
and editing steps.

\noindent \textbf{Limited telemetry signal in TT-Java.}
\label{sec:rq1-ttc4}
The TT-Java value faults change returned values but produce little observable
change in the collected logs, traces, or metrics. Repeated paired collections
also showed no separable telemetry difference for this fault class. We therefore
treat the test-based syntactic and semantic correctness check as the primary
TT-Java repair outcome and interpret telemetry replay for TT-Java cautiously.
\textsc{ORCA} emits patches for $98$ of $100$ cases and obtains File hits in
$94$, but only $32$ patches both pass patch validity and compile; among these,
$30$ pass the tests. The second-highest test-pass count is mini-SWE-agent's
$17$. These results indicate that, for TT-Java, File hit is high but many
generated patches do not reach the syntactic and semantic correctness check
because they fail patch validity or compilation.

\subsubsection{Synthetic configuration subset (225 cases)}
\label{sec:res-config}
The right half of Table~\ref{tab:syn200} reports results for the synthetic configuration subset. \textsc{ORCA}'s repair graph agents produce
$213$ patches that pass patch validity ($95\,\%$), compared with $177$
for mini-SWE-agent, $30$ for ReAct, and $40$--$41$ for Direct and One-Shot.
\textsc{ORCA} also obtains $131$ strict telemetry-replay cases ($58\,\%$) and
$210$ mitigated telemetry-replay cases ($93\,\%$). The highest corresponding
baseline counts are from mini-SWE-agent, with $110$ strict and $174$
mitigated telemetry-replay cases.

Direct, One-Shot, and ReAct obtain high File hit counts
($207$--$218$ of $225$ cases), but only $30$--$41$ of their patches pass patch
validity. Thus, many baselines identify the affected configuration file without
producing an applicable deployment-configuration patch. The fault signature
often provides enough service or file context for file-level localization, but
the generated diffs frequently fail before telemetry replay can evaluate them. Agentless and AutoCodeRover are not evaluated on this subset because their
released pipelines target Python code rather than deployment-configuration
repair. \textsc{ORCA} reaches a $9.9$k-token median cost
(mean $35.5$k, increased by the $20\,\%$ of cases routed to the exploration
agent), with about one-third of mini-SWE-agent's mean token cost while
producing more mitigated telemetry-replay cases.

\subsubsection{Real-incident subset (150 cases)}
\label{sec:res-real}

The real-incident subset contains the $56$ reproduced real incidents
and the $94$ mutation variants described in Section~\ref{sec:datasets}. Figure~\ref{fig:real150} reports file-level localization, patch validity,
telemetry replay, and token cost for this subset.

\definecolor{cDirect}{HTML}{8C8C8C}
\definecolor{cOne}{HTML}{5B8FB9}
\definecolor{cAg}{HTML}{4CA64C}
\definecolor{cRe}{HTML}{E89A3C}
\definecolor{cMini}{HTML}{9B6FC4}
\definecolor{cACR}{HTML}{2E5FA3}
\definecolor{cORCA}{HTML}{D1352B}

\pgfplotsset{
  realbar/.style={
    ybar,
    /pgf/bar width=6.1pt,
    ymin=0,
    xmin=0.25,
    xmax=7.75,
    enlarge x limits=false,
    scale only axis,
    xtick=\empty,
    tick label style={font=\footnotesize},
    ylabel style={font=\footnotesize},
    point meta=explicit symbolic,
    nodes near coords,
    nodes near coords style={font=\tiny,rotate=90,anchor=west,inner sep=1pt},
    every axis plot/.append style={/pgf/bar shift=0pt},
  },
}

\begin{figure*}[t]
\centering
{\footnotesize
\legbox{fill=cDirect}\,Direct\enspace
\legbox{fill=cOne}\,One-Shot\enspace
\legbox{fill=cAg}\,Agentless\enspace
\legbox{fill=cRe}\,ReAct\enspace
\legbox{fill=cMini}\,mini-SWE\enspace
\legbox{fill=cACR}\,AutoCodeRover\enspace
\legbox{fill=cORCA}\,ORCA (ours)
}\\[3pt]

\begin{tabular}{@{}l@{\hspace{-0.006\textwidth}}c@{\hspace{0.002\textwidth}}c@{}}
\begin{minipage}[t]{0.45\textwidth}
\vspace{0pt}
\begin{tikzpicture}[baseline=(current bounding box.north)]
\begin{groupplot}[
  realbar,
  group style={group size=3 by 1,horizontal sep=0.58cm},
  width=1.82cm,
  height=3.45cm
]
\nextgroupplot[ymax=104,ylabel={\# cases}]
\addplot[fill=cDirect]coordinates{(1,6)[6 (4\%)]};
\addplot[fill=cOne]coordinates{(2,5)[5 (3\%)]};
\addplot[fill=cAg]coordinates{(3,1)[1]};
\addplot[fill=cRe]coordinates{(4,19)[19 (13\%)]};
\addplot[fill=cMini]coordinates{(5,58)[58 (39\%)]};
\addplot[fill=cACR]coordinates{(6,1)[1]};
\addplot[fill=cORCA]coordinates{(7,49)[49 (33\%)]};

\nextgroupplot[ymax=205]
\addplot[fill=cDirect]coordinates{(1,27)[27]};
\addplot[fill=cOne]coordinates{(2,29)[29]};
\addplot[fill=cAg]coordinates{(3,74)[74 (49\%)]};
\addplot[fill=cRe]coordinates{(4,26)[26]};
\addplot[fill=cMini]coordinates{(5,74)[74 (49\%)]};
\addplot[fill=cACR]coordinates{(6,19)[19]};
\addplot[fill=cORCA]coordinates{(7,113)[113 (75\%)]};

\nextgroupplot[ymax=86]
\addplot[fill=cDirect,fill opacity=0.38]coordinates{(1,1)[1]};
\addplot[fill=cOne,fill opacity=0.38]coordinates{(2,2)[2]};
\addplot[fill=cRe,fill opacity=0.38]coordinates{(4,1)[1]};
\addplot[fill=cMini,fill opacity=0.38]coordinates{(5,23)[23 (15\%)]};
\addplot[fill=cACR,fill opacity=0.38]coordinates{(6,1)[1]};
\addplot[fill=cORCA,fill opacity=0.38]coordinates{(7,45)[45 (30\%)]};
\addplot[fill=cMini,forget plot]coordinates{(5,7)[7]};
\addplot[fill=cORCA,forget plot]coordinates{(7,16)[16]};
\end{groupplot}
\end{tikzpicture}
\label{fig:real150-eff}
\end{minipage}
&
\begin{minipage}[t]{0.20\textwidth}
\centering
\vspace{0pt}
\begin{tikzpicture}[baseline=(current bounding box.north)]
\begin{axis}[
  realbar,
  /pgf/bar width=6.4pt,
  ymode=log,
  log origin=infty,
  ymin=4,
  ymax=2000,
  ytick={10,100,1000},
  width=1.85cm,
  height=3.45cm,
  ylabel={tokens/case (k)}
]
\addplot[fill=cDirect]coordinates{(1,8)[8]};
\addplot[fill=cOne]coordinates{(2,7)[7]};
\addplot[fill=cAg]coordinates{(3,14)[14]};
\addplot[fill=cRe]coordinates{(4,31)[31]};
\addplot[fill=cMini]coordinates{(5,640)[640]};
\addplot[fill=cACR]coordinates{(6,109)[109]};
\addplot[fill=cORCA]coordinates{(7,26)[26]};
\end{axis}
\end{tikzpicture}
\label{fig:real150-cost}
\end{minipage}
&
\begin{minipage}[t]{0.32\textwidth}
\centering
\vspace{0pt}
\includegraphics[width=\linewidth]{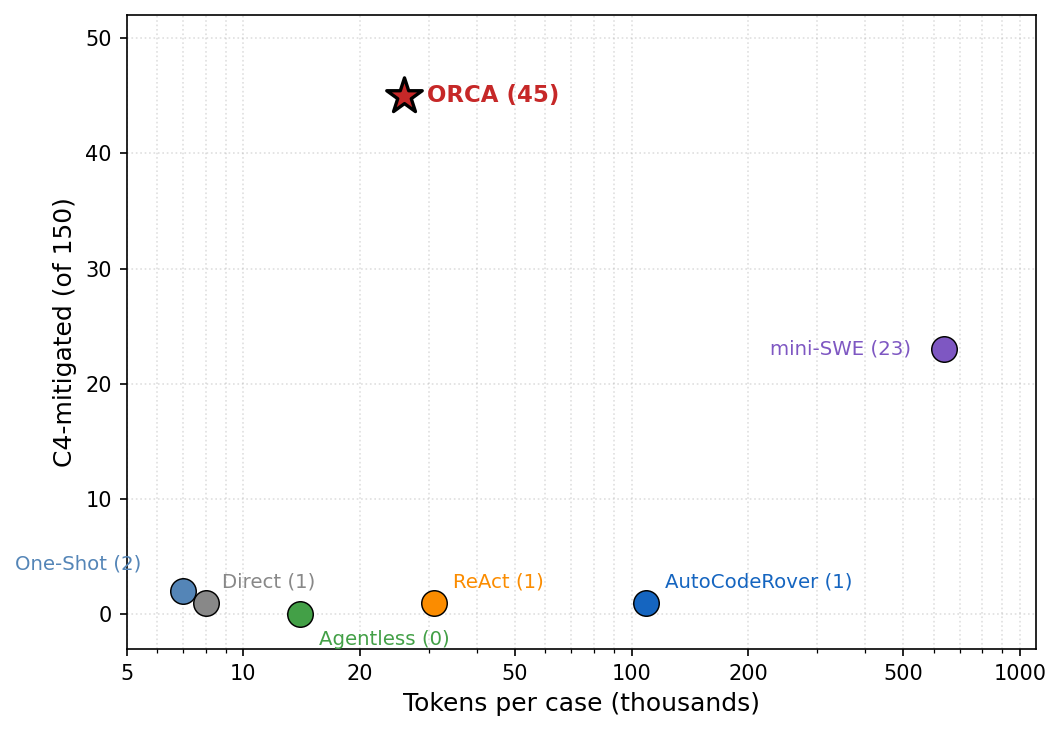}
\label{fig:real150-tradeoff}
\end{minipage}
\\[-15pt]
\begin{minipage}[t]{0.45\textwidth}
\hspace*{0.05\textwidth}{\footnotesize (a) File hit, C1 patch validity, and C4 telemetry replay outcomes}
\end{minipage}
&
\begin{minipage}[t]{0.20\textwidth}
\centering
\hspace*{0.15\textwidth}{\footnotesize\mbox{(b) Tokens/case}}
\end{minipage}
&
\begin{minipage}[t]{0.32\textwidth}
\centering
{\footnotesize (c) Mitigated telemetry replay vs.\ tokens per case}
\end{minipage}
\end{tabular}
\vspace{-5pt}
\caption{Results of seven systems on the real-incident subset ($150$ cases).
For telemetry replay outcomes, faded bars denote mitigated telemetry replay and solid
bars denote strict telemetry replay.}
\label{fig:real150}
\end{figure*}
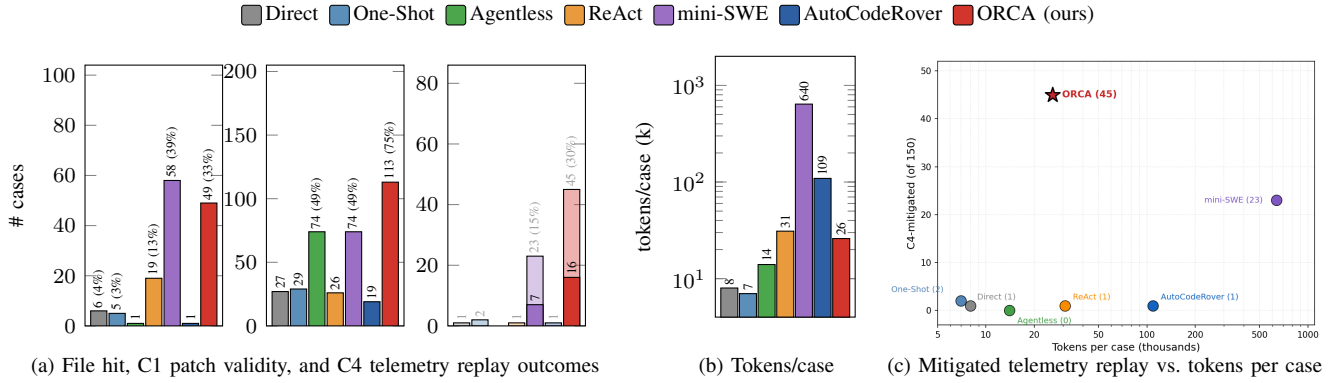

\textsc{ORCA} produces $113$ patches that pass patch validity, $45$
telemetry-replay mitigations, and $16$ strict telemetry-replay cases. The
corresponding counts for mini-SWE-agent are $74$, $23$, and $7$. Relative to
mini-SWE-agent, \textsc{ORCA} obtains $1.53$ times as many patches that pass
patch validity, $1.96$ times as many telemetry-replay mitigations, and $2.29$
times as many strict telemetry-replay cases. Its mean token cost is also lower:
$26$k tokens per case, compared with $640$k for mini-SWE-agent. This comparison
shows that, on the real-incident subset, \textsc{ORCA} improves the observed
verification outcomes while using substantially fewer tokens per case.

Despite having fewer File hits than mini-SWE-agent
($49$ versus $58$), \textsc{ORCA} produces more applicable patches and
more telemetry-replay mitigations. One reason is that File hit does not
capture alternate-location repairs: for example, Section~\ref{sec:res-c4}
shows a configuration patch that edits a different file from the upstream
repair while disabling the same faulty instrumentation. More generally,
\textsc{ORCA}'s repair graph agents convert localized candidates into
patches that pass patch validity more often than the evaluated baselines. Direct, One-Shot, and ReAct produce
few telemetry-replay mitigations. Agentless and AutoCodeRover are limited by
their Python-specific localization and repair components; the
real-incident subset also includes non-Python source files and
deployment-configuration repairs, so these systems cover only part of the
subset's repair space.

To assess whether telemetry replay check (C4) reflects meaningful fault mitigation rather than only
changes to telemetry emission, we manually inspected the $45$ \textsc{ORCA}
patches that satisfied strict or mitigated telemetry replay on the real-incident
subset. We compared each generated diff with the issue title and body, the
upstream repair, and the replayed fault signature, and checked whether the edit
corrects, removes, or disables the fault-related code path,
deployment-configuration key, telemetry key, or instrumentation setting. This
inspection finds that $31$ of the $45$ telemetry-replay mitigations are
fault-addressing patches, including $11$ File hit cases and $20$ File miss
cases. The remaining $14$ cases reduce the replayed fault signature but are not
counted as manually confirmed repairs. This result supports using C4 as
evidence that a patch can address or partially mitigate the observed fault,
while C4 remains a diagnostic signal rather than a complete correctness oracle.

\subsection{RQ2: Sensitivity to the LLM backbone}
\label{sec:rq2}

RQ2 evaluates model sensitivity on the 150-case real-incident subset. We compare
seven LLMs from four providers: DeepSeek-V4-pro, DeepSeek-Flash,
Claude-Opus-4.8, GPT-5.1, GPT-5-mini, GPT-4o-mini, and
Qwen3-Coder-30B~\cite{deepseek_v4_pro,deepseek_flash,claude_opus_4_8,
openai_gpt51,openai_gpt5mini,openai_gpt4omini,qwen3_coder_30b}. We keep the
\textsc{ORCA} pipeline, prompts, input data, temperature, and reasoning effort
fixed, using temperature $0$ and low reasoning effort for all models, and vary
only the LLM models(Table~\ref{tab:rq2}). Across these models, File hit
ranges from $41$ to $57$ cases, patch validity from $113$ to $127$ cases, and
telemetry-replay mitigations from $37$ to $45$ cases. Claude-Opus-4.8 obtains
the highest File hit and patch-validity counts, while DeepSeek-V4-pro and
GPT-4o-mini obtain the highest telemetry-replay mitigation count.

The larger variation is in token cost. Mean tokens per case range from
$6.9$k with GPT-4o-mini to $26.4$k with DeepSeek-V4-pro, a $3.8$ times
difference. In this experiment, changing the LLM backbone changes token cost more
than the reported File hit, patch-validity, and telemetry-replay mitigation
counts. This pattern is consistent with \textsc{ORCA}'s telemetry-localized repair
setting. Telemetry-Based Fault Localization identifies candidate code and
configuration locations from telemetry-derived evidence, and the repair graph
agents package these locations with local code or configuration context and
repository-structure relations. The LLM backbone is therefore asked to synthesize
an exact-match edit within a constrained repair context, rather than to infer
repair locations primarily from the issue title and body and the repository.
Model choice still affects cost and some outcomes, but on this subset the
evaluated models produce similar File hit, patch-validity, and
telemetry-replay mitigation counts under the same \textsc{ORCA} pipeline.

\begin{table}[t]
\renewcommand{\arraystretch}{1.2}
\caption{LLM-backbone sensitivity on the real-incident subset.}
\label{tab:rq2}
\centering\footnotesize
\vspace{-5pt}
\begin{tabular}{@{}lrrrr@{}}
\toprule
\textbf{Model} & \textbf{File hit/150} & \textbf{C1/150} & \textbf{C4-mit.} & \textbf{Tokens} \\
\midrule
Claude-Opus-4.8 & \textbf{57} & \textbf{127} & 43 & 14.5k \\
GPT-5.1 & 52 & 121 & 43 & 11.1k \\
DeepSeek-V4-pro & 49 & 113 & \textbf{45} & 26.4k \\
Qwen3-Coder-30B & 52 & 117 & 43 & 10.5k \\
DeepSeek-Flash & 49 & 117 & 37 & 20.1k \\
GPT-5-mini & 41 & 116 & 44 & 14.9k \\
GPT-4o-mini & 41 & 120 & \textbf{45} & \textbf{6.9k} \\
\bottomrule
\end{tabular}
\end{table}

\subsection{RQ3: Effect of the Exploration-agent run count $N$}
\label{sec:rq3}

RQ3 evaluates the Exploration-agent run count on real-incident cases routed to
this stage. Of the $150$ cases in the real-incident subset, $40$ are routed to
the Exploration agent. On these $40$ cases, we
compare $N\in\{1,3,5\}$ and combine multiple runs using the file-level voting rule of Section~\ref{sec:hybrid-exploration}, keeping the model
(GPT-5-mini), prompts, and inputs fixed.

Increasing $N$ improves several repair outcomes. File hit rises from $4$
at $N{=}1$ to $10$ at $N{=}3$ and $11$ at $N{=}5$, while patch validity
rises from $28$ to $35$ at $N{=}3$ and remains above the $N{=}1$ setting at
$N{=}5$. Telemetry-replay mitigations remain similar at $7$--$8$ cases.
The improvement comes at higher aggregate cost over the $40$ routed cases:
token use increases from $1.6$M to $4.5$M tokens, or $2.81$ times. We therefore use $N{=}1$ elsewhere in
the evaluation for efficiency.

\subsection{RQ4: Sensitivity to LLM temperature}
\label{sec:rq4}
RQ4 evaluates sensitivity to LLM temperature on the 150-case
real-incident subset. We keep the \textsc{ORCA} pipeline, input data,
and LLM backbone fixed, and vary only the temperature from $T{=}0$ to $0.9$
using gpt-4o-mini, which supports the temperature range used in
this experiment. For each temperature, we use $R{=}3$ random seeds for generation
metrics; telemetry replay is measured once per temperature.

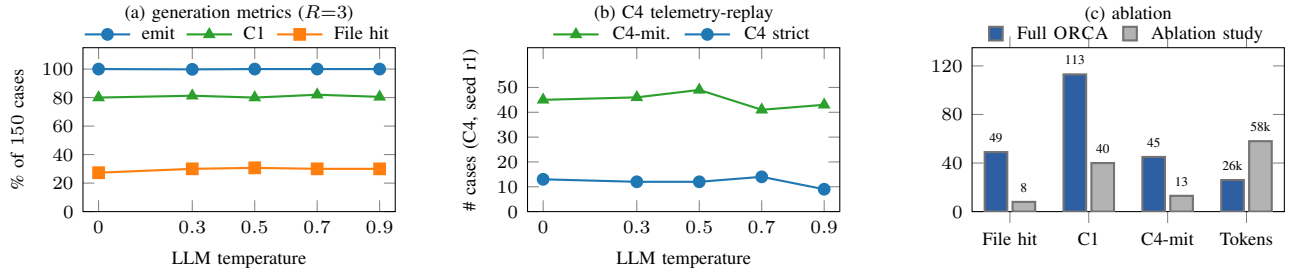
\begin{figure*}[t]
\centering
\definecolor{cFull}{HTML}{2E5FA3}
\definecolor{cBlind}{HTML}{B3B3B3}

\begin{tikzpicture}
\begin{groupplot}[
  group style={group size=3 by 1, horizontal sep=1.75cm},
  width=0.315\textwidth,
  height=3.75cm,
  tick label style={font=\scriptsize},
  label style={font=\scriptsize},
  title style={font=\scriptsize},
  legend cell align={left},
  legend style={font=\scriptsize, draw=none, fill=none},
  every axis plot/.append style={thick, mark size=2pt},
]

\nextgroupplot[
  title={(a) generation metrics ($R{=}3$)},
  xlabel={LLM temperature},
  ylabel={\% of $150$ cases},
  xtick={0,0.3,0.5,0.7,0.9},
  xmin=-0.05,
  xmax=0.95,
  ymin=0,
  ymax=115,
  ytick={0,20,40,60,80,100},
  legend style={
    at={(0.5,1.2)},
    anchor=north,
    draw=none,
    fill=none,
    font=\scriptsize,
    legend columns=3,
    column sep=2pt,
    /tikz/every even column/.append style={column sep=2pt}
  },
]
\addplot[mplblue, mark=*] coordinates {(0,100)(0.3,99.8)(0.5,100)(0.7,100)(0.9,100)};
\addlegendentry{emit}
\addplot[mplgreen, mark=triangle*] coordinates {(0,80.0)(0.3,81.3)(0.5,80.0)(0.7,82.0)(0.9,80.5)};
\addlegendentry{C1}
\addplot[mplorange, mark=square*] coordinates {(0,27.3)(0.3,30.0)(0.5,30.7)(0.7,30.0)(0.9,30.0)};
\addlegendentry{File hit}

\nextgroupplot[
  title={(b) C4 telemetry-replay},
  xlabel={LLM temperature},
  ylabel={\# cases (C4, seed r1)},
  xtick={0,0.3,0.5,0.7,0.9},
  xmin=-0.05,
  xmax=0.95,
  ymin=0,
  ymax=66,
  ytick={0,10,20,30,40,50},
  legend style={
    at={(0.5,1.2)},
    anchor=north,
    draw=none,
    fill=none,
    font=\scriptsize,
    legend columns=2,
    column sep=3pt,
    /tikz/every even column/.append style={column sep=3pt}
  },
]
\addplot[mplgreen, mark=triangle*] coordinates {(0,45)(0.3,46)(0.5,49)(0.7,41)(0.9,43)};
\addlegendentry{C4-mit.}
\addplot[mplblue, mark=*] coordinates {(0,13)(0.3,12)(0.5,12)(0.7,14)(0.9,9)};
\addlegendentry{C4 strict}

\nextgroupplot[
  title={(c) ablation},
  ybar,
  /pgf/bar width=8.5pt,
  ymin=0,
  ymax=135,
  ytick={0,40,80,120},
  symbolic x coords={g,c1,c4,t},
  xtick=data,
  xticklabels={File hit,C1,C4-mit,Tokens},
  x tick label style={font=\scriptsize},
  enlarge x limits=0.16,
  point meta=explicit symbolic,
  nodes near coords,
  nodes near coords style={font=\tiny},
  every axis plot/.append style={draw=black!55},
  legend style={
    at={(0.5,1.2)},
    anchor=north,
    draw=none,
    fill=none,
    font=\scriptsize,
    legend columns=2,
    column sep=3pt,
    /tikz/every even column/.append style={column sep=3pt}
  },
  legend image code/.code={
  \draw[#1, draw=black!55, fill=#1]
    (0cm,-0.08cm) rectangle (0.18cm,0.10cm);
},
]
\addplot[fill=cFull] coordinates {
  (g,49)[49]
  (c1,113)[113]
  (c4,45)[45]
  (t,26)[26k]
};

\addlegendentry{Full \textsc{ORCA}}

\addplot[fill=cBlind] coordinates {
  (g,8)[8]
  (c1,40)[40]
  (c4,13)[13]
  (t,58)[58k]
};
\addlegendentry{Ablation study}
\end{groupplot}
\end{tikzpicture}
\vspace{-12pt}
\caption{ Effect of LLM temperature on \textbf{(a)} generation metrics and \textbf{(b)} telemetry-replay check. \textbf{(c)} Ablation of the fault signature and issue context.}
\label{fig:rq4_ablation}
\end{figure*}

Across the temperature range, patch generation remains near $100\,\%$.
Patch validity stays within $120$--$123$ cases, File hit within
$41$--$46$ cases, and telemetry-replay mitigations within $41$--$49$ cases
(Figure~\ref{fig:rq4_ablation}(a)--(b)). Mean token cost remains between $6.7$k and $7.0$k
tokens per case. Strict telemetry-replay counts range from $9$ to $14$ cases,
with the lowest value at $T{=}0.9$. Overall, the measured outcomes vary
within a narrow range across the evaluated temperature settings. We use
$T{=}0$ in the other experiments to keep generation deterministic.

\subsection{Ablation: contribution of the fault signature and issue context}
\label{sec:ablation-telemetry}

This ablation study isolates two sources of repair context in the real-incident
pipeline: the fault signature derived from paired failure and reference
telemetry, and the sanitized issue context derived from the issue title and
body. We rerun the 150-case real-incident subset with both inputs removed from
the full pipeline, including localization and patch generation, keep the rest
of the pipeline configuration fixed, and compare this variant with full
\textsc{ORCA} (Figure~\ref{fig:rq4_ablation}(c)).
Removing these inputs reduces all reported repair outcomes: File hit
decreases from $49$ to $8$, patch validity from $113$ to $40$, and
telemetry-replay mitigations from $45$ to $13$. Mean token cost increases from $26$k to $58$k tokens per case, consistent with additional
workspace exploration before patch generation. These results indicate that the
distilled fault signature and sanitized issue context provide useful repair
context for both localization and patch construction.

\subsection{Case study}
\label{sec:case-studies}
\noindent
\textbf{Telemetry mitigation with a File miss.}
\label{sec:res-c4}
Telemetry replay can also identify mitigations that are not counted by the
File hit metric. Issue \#1468 of
\textit{opentelemetry-demo}~\cite{oteldemo1468} is one such case
(Figure~\ref{fig:case1468}). The frontend's filesystem
auto-instrumentation emits many additional \textit{fs} spans,
which dominate the trace stream. The upstream repair disables this
instrumentation in the frontend Node SDK code, whereas \textsc{ORCA}'s
configuration repair graph agent disables the same instrumentation through
the documented deployment environment variable. The two patches therefore
edit different files, but they target the same telemetry source. Under
telemetry replay, the \textit{fs}-span fault signature decreases from
$1.0$ to $0.0$, while the expected-behavior score remains $1.0$. Because the \textsc{ORCA} patch edits a different file from the upstream
repair, it is counted as a File miss; telemetry replay, however, records the
reduction in the fault signature as a mitigation.

\begin{figure*}[t]
\centering
\lstdefinelanguage{wtdiff}{morecomment=[f][\color{green!48!black}]{+},morecomment=[f][\color{red!68!black}]{-}}
\lstdefinestyle{wt}{language=wtdiff,basicstyle=\ttfamily\scriptsize,breaklines=true,frame=single,framesep=3pt,xleftmargin=3pt,xrightmargin=3pt,columns=fullflexible,keepspaces=true,aboveskip=2pt,belowskip=0pt}
\begin{minipage}[t]{0.49\textwidth}
\footnotesize\textit{Upstream repair} --- \textit{src/frontend/\ldots/Instrumentation.js}\par\smallskip
\begin{lstlisting}[style=wt]
  '@opentelemetry/instrumentation-fs': {
-   requireParentSpan: true,
+   enabled: false,
\end{lstlisting}
\end{minipage}\hfill
\begin{minipage}[t]{0.49\textwidth}
\footnotesize\textbf{\textsc{ORCA}} --- \textit{docker-compose.yml}\par\smallskip
\begin{lstlisting}[style=wt]
  environment:
    - WEB_OTEL_SERVICE_NAME=frontend-web
+   - OTEL_INSTRUMENTATION_FS_ENABLED=false
\end{lstlisting}
\end{minipage}
\caption{Incident \#1468 of \textit{opentelemetry-demo}: \textsc{ORCA}
changes deployment configuration and reduces the replayed \textit{fs}-span
fault signature without a File hit.}
\label{fig:case1468}
\end{figure*}

\section{Discussion}
\label{sec:discussion}
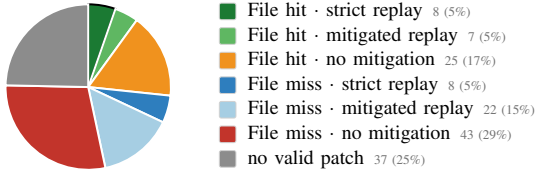
\begin{figure}[t]
\centering
\definecolor{cF1}{HTML}{1B7A33}\definecolor{cF2}{HTML}{5FB762}\definecolor{cF3}{HTML}{F0921E}%
\definecolor{cF4}{HTML}{2E7EBE}\definecolor{cF5}{HTML}{A6CEE3}\definecolor{cF6}{HTML}{C2342B}\definecolor{cF7}{HTML}{8C8C8C}%
\newcommand{\taxslice}[4]{\fill[#3,draw=#4,line width=1.2pt] (0,0) -- (#1:\R) arc(#1:#2:\R) -- cycle;}%
\resizebox{0.80\columnwidth}{!}{%
\begin{tikzpicture}
\def\R{1.75}

\taxslice{90}{70.8}{cF1}{black}
\taxslice{70.8}{54}{cF2}{white}
\taxslice{54}{-6}{cF3}{white}
\taxslice{-6}{-25.2}{cF4}{white}
\taxslice{-25.2}{-78}{cF5}{white}
\taxslice{-78}{-181.2}{cF6}{white}
\taxslice{-181.2}{-270}{cF7}{white}

\foreach \i/\c/\lab/\val in {
  0/cF1/{File hit $\cdot$ strict replay}/{8 (5\%)},
  1/cF2/{File hit $\cdot$ mitigated replay}/{7 (5\%)},
  2/cF3/{File hit $\cdot$ no mitigation}/{25 (17\%)},
  3/cF4/{File miss $\cdot$ strict replay}/{8 (5\%)},
  4/cF5/{File miss $\cdot$ mitigated replay}/{22 (15\%)},
  5/cF6/{File miss $\cdot$ no mitigation}/{43 (29\%)},
  6/cF7/{no valid patch}/{37 (25\%)}
}{
\pgfmathsetmacro\yy{\R-0.15-\i*0.52}  \node[
    draw=black!18,
    fill=\c,
    rounded corners=1.2pt,
    minimum width=10pt,
    minimum height=10pt,
    inner sep=0pt
  ] at (2.95,\yy){};
  \node[anchor=west,font=\large] at (3.25,\yy)
    {\lab\enspace{\footnotesize\textcolor{black!55}{\val}}};
}
\end{tikzpicture}}
\vspace{-8pt}
\caption{Outcome breakdown for \textsc{ORCA} on the real-incident subset}
\label{fig:ftax}
\end{figure}

Figure~\ref{fig:ftax} partitions the $150$ real-incident cases by patch
validity, File hit/File miss status, and telemetry-replay outcome. The
breakdown clarifies how file-level localization and telemetry replay provide
different evidence about generated patches.

\noindent\textbf{Metric interpretation.}
File hit is useful for interpreting whether a patch overlaps the upstream
repair files, but not exactly a correctness measure. \textsc{ORCA} obtains File
hits in $49$ cases; $40$ of these patches pass patch validity, and $15$ of the
valid File hit patches satisfy strict or mitigated telemetry replay. Conversely,
$30$ valid File miss patches satisfy strict or mitigated telemetry replay, and
the manual inspection reported in RQ1 labels $20$ of them as fault-addressing
patches. These outcomes indicate that File hit and telemetry replay should be
reported separately: the former measures file-level overlap with the upstream
repair, while the latter measures reduction of the fault-signature telemetry.

\noindent\textbf{Failure modes.}
The two largest remaining categories are File miss patches with no telemetry
mitigation ($43$ cases, $29\,\%$) and no valid patch ($37$ cases, $25\,\%$).
The first category points to localization gaps in telemetry-sparse or
multi-language cases; the second points to patch-construction failures,
especially generated patches that fail the validity check.

\noindent\textbf{Internal threats to validity.} Telemetry replay can credit patches that alter telemetry emission rather than repair the root cause, and can be inconclusive when a fault lacks a detectable telemetry signature. We mitigate this risk by reporting both strict and mitigated telemetry replay, separating File miss telemetry-replay outcomes from File hits, and manually inspecting the telemetry-replay mitigations in RQ1. Sensitivity to LLM backbone, Exploration-agent run count, and temperature is reported in RQ2--RQ4.

\noindent\textbf{External threats to validity.}
The benchmark draws on four open-source microservice projects and covers
both code-level and deployment-configuration faults. The conclusions may not transfer unchanged to systems with different deployment substrates, proprietary
operational practices, or telemetry coverage. The ablation further shows that removing the fault signature and
sanitized issue context substantially reduces repair outcomes on this benchmark;
this dependence may differ when the issue title and body are absent or
uninformative.
Of the $150$ real-incident subset cases, $94$ are mutation variants derived
from the collected real incidents. The aggregate results should therefore be interpreted as combining reproduced incidents with controlled variants, rather than as
$150$ independently observed incidents. Finally,
the telemetry comes from reproduced workload replay rather than production
traffic, and replay-environment drift may affect telemetry-replay outcomes.

\section{Conclusion}
\label{sec:concl}

This paper presents \textsc{ORCA}, an observability-grounded repair approach for microservice incidents. \textsc{ORCA} derives a fault signature and candidate code and deployment-configuration locations from paired failure and reference telemetry, generates unified-diff patch candidates through repair graph agents and exploration, and evaluates them with TGPV checks. Across a 575-case benchmark spanning synthetic code faults, synthetic configuration faults, and real microservice incidents, \textsc{ORCA} produces more patches that pass patch validity and more telemetry-replay mitigations than the evaluated baselines, while using substantially fewer tokens than the strongest agentic baseline on the real-incident subset. These results show that operational telemetry can be transformed from diagnostic evidence into actionable repair context: paired telemetry supports repair-oriented localization, while repair graph
agents convert localized code and configuration evidence into constrained
patch-generation context for the LLM. Telemetry-grounded verification then
exposes repair outcomes that issue- or test-only evaluation would miss.

\balance

\bibliographystyle{IEEEtran}
\bibliography{refs}

\end{document}